\documentclass[vecphys]{svmult}

\usepackage{graphicx}        

\newcommand{\comment}[1]{}

\begin{document}

\title*{Stellar Velocity Distribution in Galactic Disks}
\author{Christian Theis\inst{1}\and
        Eduard Vorobyov\inst{2}}
\institute{Institute of Astronomy, University of Vienna, 
          T\"urkenschanzstr.\ 17, 1180 Vienna, Austria,
          \texttt{theis@astro.univie.ac.at}
\and Institute for Computational Astrophysics, Saint Mary's University,
          Halifax, B3H 3C3, Canada,
          \texttt{vorobyov@ap.smu.ca}
\and Institute of Physics, South Federal University, Stachki 194,
          Rostov-on-Don, 344090, Russia}
%
%
\maketitle

\begin{abstract}
  We present numerical studies of the properties of the stellar
  velocity distribution in galactic disks which have developed a
  saturated, two-armed spiral structure.  In previous papers we used
  the Boltzmann moment equations (BME) up to second order for our
  studies of the velocity structure in self-gravitating stellar disks.
  A key assumption of our BME approach is the zero-heat flux
  approximation, i.e.\ the neglection of third order velocity terms.
  We tested this assumption by performing test particle simulations
  for stars in a disk galaxy subject to a rotating spiral
  perturbation. As a result we corroborated qualitatively the complex
  velocity structure found in the BME approach. It turned out that an
  equilibrium configuration in velocity space is only slowly
  established on a typical timescale of 5 Gyrs or more. Since many
  dynamical processes in galaxies (like the growth of spirals or bars)
  act on shorter timescales, pure equilibrium models might not be
  fully appropriate for a detailed comparison with observations like
  the local Galactic velocity distribution.  Third order velocity
  moments were typically small and uncorrelated over almost all of the
  disk with the exception of the 1:4 resonance region (UHR).
  Near the UHR (normalized) fourth and fifth order velocity
  moments are still of the same order as the second and third order
  terms. Thus, at the UHR higher order terms are not negligible.
\end{abstract}

\section{Introduction}
\label{sectheis:intro}

It is long known that the velocity
distribution of stars in galactic disks is non-isotropic
\cite{kobold90}.  Kapteyn and Eddington
invoked a superposition of (isotropic) stellar streams with different
mean velocities, by this creating an anisotropic velocity distribution
(\cite{kapteyn05}, \cite{eddington06}).  However, an alternative
interpretation by Schwarzschild became the general framework for
describing the local velocity distribution of stars of equal age
\cite{schwarzschild07}.  The Schwarzschild distribution is based on a
single but anisotropic ellipsoidal distribution function. It is
characterized by a Gaussian distribution in all three directions $U$
(radial $r$), $V$ (tangential $\phi$) and $W$ (vertical $z$) in
velocity space, but with different velocity dispersions $\sigma_{rr}$,
$\sigma_{\phi\phi}$, and $\sigma_{zz}$. In general, the velocity
distribution is described by a velocity dispersion tensor
\begin{equation}
  \sigma^2_{ij} \equiv \overline{(v_i - \bar{v}_i)(v_j - \bar{v}_j)},
  \label{mixed_vd}
\end{equation}
where $i$ and $j$ denote the different coordinate directions and $v_i$
gives the corresponding velocity. The bar denotes local averaging over
velocity space.  The principal axes of this tensor form an imaginary
surface that is called the velocity ellipsoid. This ellipsoid is
characterized by its anisotropy, measured by the ratio of the velocity
dispersions along the principal axes
and its orientation.  Because the principal axes need not to be
aligned with the coordinate axes, the vertex deviation $l_{\rm v}$
(which is defined as the angle between the direction from the Sun to
the Galactic centre and the direction of the major principal axis of
the velocity ellipsoid) needs not to vanish.

In the case of stationary, axisymmetric systems and appropriate
distribution functions (DF) the velocity ellipsoid is aligned with the
coordinate axes, i.e.\ $l_{\rm v} = 0$.
However, non-vanishing vertex deviations were found in the solar
vicinity in many studies (\cite{wielen74}, \cite{mayor72},
\cite{ratnatunga97}). These vertex deviations could be explained by
spiral structures (\cite{mayor70}, \cite{yuan71}), by this supporting
the importance of non-axisymmetric gravitational potentials.

A major step in analysing the local stellar velocity space was the
astrometric satellite mission Hipparcos by ESA \cite{esa97}. Earlier
results concerning the general behaviour of the velocity ellipsoid,
e.g.\ the dependence of the dispersions or the vertex deviations on
$B-V$ were corroborated (\cite{dehnen98b}, \cite{bienayme99},
\cite{hogg05}). Moreover, the more numerous known proper motions
allowed for a detailed mapping of the velocity space in the solar
vicinity.  Studies by Dehnen or by Alcob\'e \& Cubarsi turned out that
the local velocity distribution of stars exhibits a rich
substructure (\cite{dehnen98a}, \cite{alcobe05}). For example, two
major peaks in the velocity distribution were found where the smaller
secondary peak is well detached by at least 30 km\,s$^{-1}$ from the
main peak (the ``u anomaly'').  Dehnen attributed this bimodality to
the perturbation exerted by the Galactic bar assuming that its outer
Lindblad resonance (OLR) is close to the Sun \cite{dehnen00}.
M\"uhlbauer \& Dehnen showed that a central bar can also explain
vertex deviations of about 10$^\circ$ 
\cite{muehlbauer03}.  

In general, a non-axisymmetric gravitational potential might lead to a
misalignment of the velocity ellipsoid.  In the case of spiral
perturbations, this conclusion was made e.g.\ by Mayor
(\cite{mayor70}, for a review see Kuijken \&
Tremaine \cite{kuijken91}) and numerically confirmed
recently by Vorobyov \& Theis (\cite{vorobyov06}, hereafter VT06).
However, it is not clear if the non-axisymmetric gravitational field
is entirely responsible for the observed vertex deviation. The
existence of moving groups of stars was also shown to produce large
vertex deviations \cite{binney98}. The situation may become even more
complicated because moving groups of stars may in turn be caused by
the non-axisymmetric gravitational field of spiral arms.  Therefore, a
detailed numerical study of the vertex deviation in spiral galaxies is
necessary.

In this paper, we present results of our analysis of the velocity
structure in disk galaxies caused by spiral perturbations. In earlier
papers (VT06, \cite{vorobyov08}) we presented numerical solutions of
the Boltzmann moment equations (BME) up to second order for flat
disks.  In Sect.\ \ref{sectheis:beads2d} we briefly describe this
method and its results with respect to the vertex deviation. A
characteristic of our BME approach is the assumption of a zero
heat-flux, i.e.\ we neglect all velocity terms of higher order than 2.
In order to test this assumption, we developed a test particle code
which measures the velocity moments up to fifth order.  This code and
first results about the importance of higher order terms are presented
in Sect.\ \ref{sectheis:testparticle}.

\section{The Boltzmann moment equations: BEADS-2D}
\label{sectheis:beads2d}

\comment{
In order to analyze the stellar velocity distribution of galactic
disks, one might either use perturbation theory or numerical models.
If non-linear effects should be considered, N-body simulations or
the Boltzmann equations can be applied. Both approaches,
however, are computationally very expensive. Therefore, there are
practically no self-consistent N-body simulations yielding
reliable results for the velocity distribution in disk galaxies.
Alternatively, restricted N-body simulations have often been used and
we will discuss such a test particle approach in Sect.\
\ref{sectheis:testparticle}.
}

We developed a numerical code (BEADS-2D) for flat stellar disks based
on the Boltzmann moment equations (BME) up to second order (VT06,
\cite{vorobyov08}). This is basically a numerical solution of the
Jeans equations. The advantage of the BME approach is twofold: first,
it allows to follow perturbations growing from a very low perturbation
amplitude up to the non-linear regime. Neither perturbation theory nor
N-body simulations can do this, for different reasons. Secondly,
observables like mean velocities or velocity dispersions can be
calculated with a very high spatial resolution all over the disk.


\begin{figure}
  \center
  \begin{minipage}{6.5cm}
    \includegraphics[height=6.5cm]{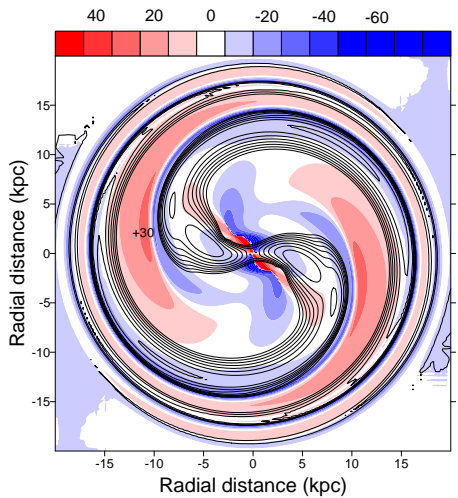}
  \end{minipage}
  \begin{minipage}{5cm}
    \caption{Positive stellar density perturbation superimposed on the
      vertex deviation map at $t=1.6$~Gyr. Positive and negative
      vertex deviations are shown in grey scale. The numbers indicate
      the maximum positive vertex deviation in the inter-arm region
      and maximum negative vertex deviation in the outer disk. The
      scale bar is in degrees.  More details about this simulation as
      well as a color figure can be found in \cite{vorobyov08}, fig.\
      9.  \label{figtheis:1}}
  \end{minipage}
\end{figure}

We studied the evolution of an initially exponential disk which
develop a two-armed spiral. Its vertex deviation shows a large spatial
variation (Fig.\ \ref{figtheis:1}). The values reach up to $90^\circ$
in the central region. A strong variation of $l_{\rm v}$ has
been found at the outer edges of the spiral structures:
$l_{\rm v}$ can vary there from $+40^\circ$ to $-40^\circ$
within only a few kpc.  The epicyclic approximation fails almost
everywhere with respect to the vertex deviation (or the Oort ratio).
For more details, see Vorobyov \& Theis \cite{vorobyov08}.

\section{Test particle simulations}
\label{sectheis:testparticle}

Our BEADS-2D code is based on the assumption of vanishing 3rd order
velocity moments. In the case of pressure-supported stellar systems like
globular clusters 3rd order terms are related to the heat flux which
is controlled by two-body relaxation. Therefore, the zero-heat flux
assumption (ZHFA) can be safely adopted on short timescales (e.g.\ a
few dynamical timescales) due to long two-body relaxation timescales
for stellar systems with more than $10^4$ stars.  However, for
rotation-supported systems like disk galaxies the situation is less
clear. Still, the two-body relaxation time is long, but now
large-scale motions might result in non-negligible higher order
moments.

In order to study the ZHFA, we performed test particle simulations for
a galactic disk. We measured the velocity moments up to fifth order at
different galactocentric radii.  Assuming a constant pattern speed
$\Omega_s$ we solved for the equations of motion of a set of test
particles in a corotating frame in polar coordinates,
\begin{eqnarray}
   \nonumber
   \ddot{R} & = & R \phi - \frac{\partial \Phi}{\partial R} 
              + 2 R \dot{\phi} \Omega_s + \Omega_s^2 R,
       \hspace*{0.8cm}
   \ddot{\phi} = -2 \frac{\dot{R}}{R} \dot{\phi}
              - \frac{1}{R^2} \frac{\partial \Phi}{\partial \phi} 
              - 2 \frac{\dot{R}}{R} \Omega_s \,.
\end{eqnarray}
$R$ and $\phi$ describe the radial and azimuthal coordinate of the
particle's trajectory, respectively.  The gravitational potential $\Phi(\vec{r},t)$
at position $\vec{r}$ and time $t$ is split into two parts. The first
part is a stationary, axisymmetric potential $\Phi_0(R)$ derived from
a given rotation curve $v_c(R)$ with the galactocentric distance $R$.
For a better comparison with other authors we chose the rotation curve
suggested by Contopoulos \& Grosb{\o}l (cf.\ eq.\ (2) in
\cite{contopoulos86} and parameters therein). The second part of the
gravitational potential was selected to be a time-dependent (rigidly
rotating) two-armed spiral perturbation also suggested in
\cite{contopoulos86}:
\begin{equation}
   \Phi(R,\phi) = f(t) \cdot A R e^{-\epsilon_s R} \cdot 
      \cos(m \ln R / \tan i_0 - 2 \phi) 
\end{equation}
For our model we adopted a two-armed ($m=2$) spiral with an amplitude $A=200 \,
\mathrm{km}^2\,\mathrm{s}^{-2}\,\mathrm{kpc}^{-1}$, an inverse scale
length $\epsilon_s=0.1 \, \mathrm{kpc}^{-1}$ and a pitch angle
$i_0=-30^\circ$.  We selected $\Omega_s = 12.5 \,
\mathrm{km}\,\mathrm{s}^{-1}\,\mathrm{kpc}^{-1}$ which puts corotation
(CR) at a radius of about 23 kpc, the 1/4-resonance (UHR) at 12 kpc
and the inner Lindblad resonance at 1.4 kpc. In order to avoid any
initial kicks the perturbation was switched-on ''adiabatically'' on a
timescale of 600 Myr.  This procedure is attributed to the function
$f(t)$ which we chose as in \cite{blitz91}.


For our statistical analysis we adopted an approach similar to that of
Blitz \& Spergel \cite{blitz91}. Initially $2 \cdot 10^6$ stars were
distributed in an exponential disk within a radial range from 4 to 40
kpc and a scale length of 10 kpc.  The initial velocities were taken
to be nearly circular (derived from the rotation curve) with a
superimposed small perturbation: the radial velocity was derived from
a Gaussian distribution with a dispersion of 20 km/s. The azimuthal
velocity perturbation was calculated from a Gaussian distribution,
taking asymmetric drift and the ratio of the radial to the azimuthal
velocity dispersions within the epicyclic approximation into account
\cite{blitz91}.

The velocity data are analysed on a polar grid ranging from 4 to 40
kpc with a radial grid spacing of 200 pc and an azimuthal spacing of
500 grid cells.  Similar to \cite{blitz91} we measured the position of
an orbit in intervals of $10^6$ years and considered its contribution
to the velocity moments of this cell. By this technique (which is
based on the ergodicity assumption), a single orbit yields many
contributions to the velocity moments, by this strongly increasing the
sample size. E.g.\ a subset of about $2.3 \cdot 10^5$ particles
hitting the radial range of 11.9 to 12.1 kpc yields about $3 \cdot
10^7$ contributions to the velocity moments in this annulus (measured
within the last 2 Gyr of the simulation).  In order to be able to detect
evolutionary effects, we sampled usually the orbits in 10 different
periods with a duration of 1 Gyr each.

\section{Results}
\label{sectheis:results}

During the simulations we calculated the cumulative velocity moments
$v_r^v v_\phi^w$ for each cell hit by a particle and for all sums
$s=v+w$ of the exponents ranging from $s=0$ up to $s=5$. From the
corresponding mean velocity moments we derived the corresponding
velocity dispersion terms including higher order terms 
$\sigma_{ijk}, \sigma_{ijkl}$ and $\sigma_{ijklm}$ with $i,j,k,l,m = r
\, \mathrm{or} \, \phi$ like
$   \sigma_{rrr} \equiv \overline{\left(v_r - u_r \right) 
                           \left(v_r - u_r \right) 
                           \left(v_r - u_r \right) }
         = \overline{v_r^3} - 3 \overline{v_r^2} u_r + 2 u_r^3
$
(with the mean radial velocity $u_r \equiv \overline{v_r}$). We
controlled our dispersion term calculation by Mathematica. 

\begin{figure}
  \center
  \includegraphics[width=6.0cm,angle=270]{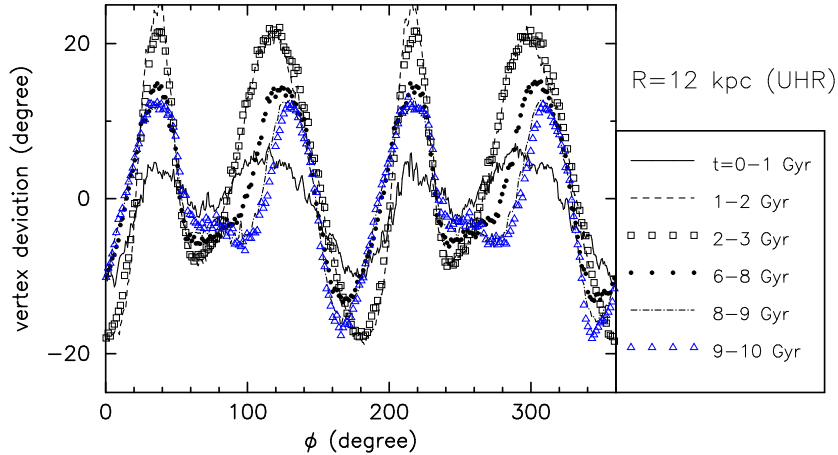}
  \caption{Azimuthal distribution of the vertex deviation
    $l_{\rm v}$ sampled at different times at a galactocentric
    distance of 12 kpc (UHR) in a annulus of 200 pc thickness.}
  \label{figtheis:2}
\end{figure}

Fig.\ \ref{figtheis:2} shows the azimuthal distribution of the vertex
deviation near the UHR (1:4 resonance) at 12 kpc. Similarly to our
BEADS-2D result a four armed-structure is visible for the vertex
deviation\footnote{Note: in the BEADS-2D model the UHR is at a radius of
  about 4.5 kpc.}. Already within the first Gyr four peaks are
visible at about the final positions (a slight shift to larger angles
can be seen for the second and the fourth peak). However, the absolute
values of the vertex deviations vary considerably in time. It is
interesting to note that the maximum values of $l_{\rm v}$
exceed the final values (at 10 Gyr) by about a factor of 2. After 6
Gyrs the final values are reached within 20\%.

This strong temporal variation is a caveat for test particle
simulations: obviously the stellar system needs a long time (by far
longer than 3 Gyr) to reach an equilibrium configuration. However, in
reality, perturbations evolve on shorter timescales: some might be
growing, others might be vanishing.  Therefore matching absolute
values derived from equilibrium configurations of test particle
simulations to observational values can be misleading, because the
stellar system might not have the time to establish an equilibrium.
The latter strongly favors either self-consistent simulations (like
the BEADS-2D approach) or non-equilibrium analyses of test particle
models.

The applicability of the BEADS-2D models depends on the validity of
the zero-heat flux approximation. Therefore, we calculated the third
(and higher) order terms all over the disk. At the UHR most of the
third order terms show a clear, well-defined fourfold structure
characteristic for the 1:4 resonance region (e.g.\ \cite{patsis06}).
Compared to the initial velocity dispersion (2nd order term) of 20
km/s, most (normalized) values of the 3rd order terms are smaller, but
not necessarily negligible.

\begin{figure}
  \center
  \includegraphics[width=7.5cm,angle=270]{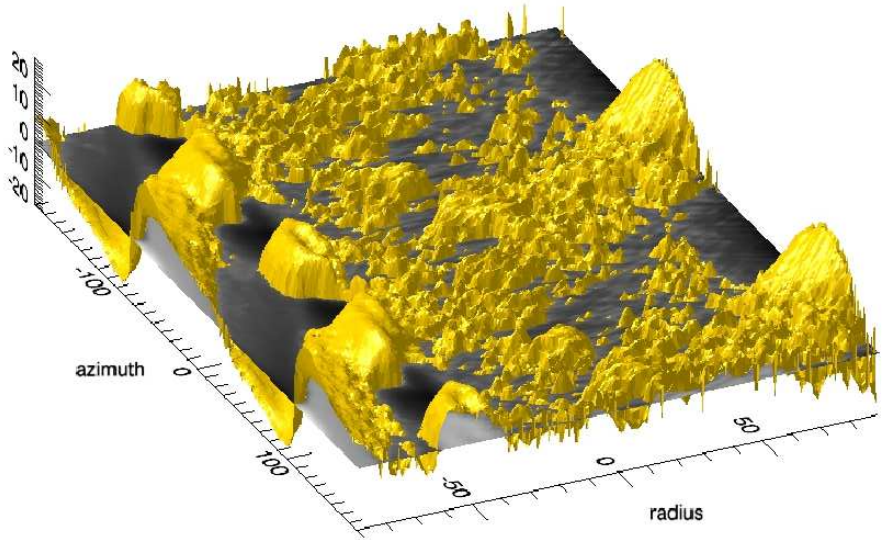}
  \caption{Distribution of the normalized third order velocity
    dispersion term $\mathrm{sign}(\sigma_{rrr}) \cdot \left(
      |\sigma_{rrr}| \right)^{1/3}$ in the galactic disks sampled in
    the period 9-10 Gyr. Both axes denote the indices of the
    corresponding polar coordinates. The radial index 0 corresponds to
    the corotation radius (23 kpc), whereas the index -90 denotes the
    inner edge of the disk (4 kpc). The UHR is at about -50.  The
    $z$-axis is given in km/s; 
    For better visualisation the data have been
    smoothed and filtered.}
  \label{figtheis:3}
\end{figure}

A view over the whole disk, however, shows that the third order terms
are especially large at the UHR and become negligible elsewhere.
E.g.\ Fig.\ \ref{figtheis:3} displays $\sigma_{rrr}$ exhibiting clearly the
four peaks associated with the 1:4 resonance. Beyond the UHR there are
no significant third order terms discernible; it just looks like
noise. Although one might speculate that some small structure can be
identified by the broader ``hills'' at corotation, it is remarkable
that corotation does not exhibit a more pronounced third order term.
The strong peaks found at the lower and higher radial range are due to
incomplete sampling: e.g.\ we did not consider stars inside 4 kpc
which gives a bias in the velocity distribution near the inner edge.



It is also interesting to look for terms higher than third order.
Near the UHR
some of them become very large. E.g.\ $\sigma_{rr\phi\phi}$
reachs (normalized) peak values of 70 km/s and never drops below
20 km/s. Though other terms are smaller, they all are of the order
of the initial velocity dispersion of the stars. Thus, there
is no strong decay in magnitude for the higher order 
velocity dispersion terms near the UHR.

%
\section{Summary}
\label{sectheis:summary}
We analysed the structure of the velocity space in galactic disks
which are subject to a spiral perturbation. In previous papers we used
the Boltzmann moment equations up to second order in order to study
the growth and saturation of spiral structure in self-gravitating
stellar disks. From this analysis we derived the properties of the
velocity ellipsoid all over the disk, namely the vertex deviations and
the Oort ratios.

A key assumption of the BME approach is the zero-heat flux
approximation, i.e.\ the neglection of third order velocity terms.  We
tested this assumption by performing test particle simulations for
stars in a disk galaxy subject to a rotating spiral perturbation. As a
result we corroborated qualitatively the complex velocity structure
found in the BME approach. It turned out that an equilibrium
configuration in velocity space is only slowly established on a
typical timescale of 5 Gyrs or more. Since many dynamical processes in
galaxies (like the growth of spirals or bars) act on shorter
timescales, pure equilibrium models might not be fully appropriate for
a detailed comparison with observations like the local Galactic
velocity distribution.  In our simulations third order velocity
moments were typically small and uncorrelated over almost all of the
disk with the exception of the 1:4 resonance region (UHR).  Near the
UHR (normalized) fourth and fifth order velocity moments are still of
the same order as the second and third
order terms. Thus, at the UHR higher order terms are not negligible.\\

\noindent 
{\bf Acknowledgements}. CT is very grateful to the organizers of the
meeting for a very interesting and inspiring conference as well as for
the financial support.  Additionally, we would like to thank Christoph
Lhotka for his advice in using Mathematica.


\end{document}